\documentstyle[sprocl,epsf]{article}

\newcommand{\mixingangle}{\theta_{\eta^\prime\hspace{-0.05em}\eta}}
\newcommand{\R}{R}

\newcommand{\QD}{Q_{\hspace{-0.1em}D}}
\newcommand{\bda}{\begin{\displaymath}\begin{array}{rl}}
\newcommand{\eda}{\end{array}\end{displaymath}}
\newcommand{\be}{\begin{equation}}
\newcommand{\ee}{\end{equation}}
\newcommand{\bdm}{\begin{displaymath}}
\newcommand{\edm}{\end{displaymath}}
\newcommand{\bea}{\begin{eqnarray}}
\newcommand{\eea}{\end{eqnarray}}

\newcommand{\no}{\nonumber \\}

\newcommand{\fs}{\; \; .}
\newcommand{\co}{\; \; ,}

\newcommand{\al}{&\!\!\!\!}
\newcommand{\eff}{{e\hspace{-0.1em}f\hspace{-0.18em}f}}

\newcommand{\QCD}{\mbox{\scriptsize Q\hspace{-0.1em}CD}}

\newcommand{\lvac}{\langle 0|\,}
\newcommand{\rvac}{\,|0\rangle}

\newcommand{\qbar}{\overline{\rule[0.42em]{0.4em}{0em}}\hspace{-0.45em}q}
\newcommand{\ubar}{\overline{\rule[0.42em]{0.4em}{0em}}\hspace{-0.5em}u}
\newcommand{\dbar}{\,\overline{\rule[0.65em]{0.4em}{0em}}\hspace{-0.6em}d}
\newcommand{\sbar}{\overline{\rule[0.42em]{0.4em}{0em}}\hspace{-0.5em}s}

\newcommand{\lsim}{\,\raisebox{-0.3em}{$\stackrel{\raisebox{-0.1em}
{$<$}}{\sim}$}\,}

\newcommand{\Mvec}{\vec{\hspace{0.04em}M}}

\begin{document}

\title{LIGHT QUARK EFFECTIVE THEORY}
\author{H. LEUTWYLER}
\address{Theory Division, CERN, CH-1211 Geneva,
Switzerland and\\
Institute for
Theoretical Physics, University of Bern, Sidlerstr. 5,\\ CH-3012 Bern,
Switzerland}

\maketitle\abstracts{The application of effective field theory methods to the
low energy structure of QCD is discussed. The analysis relies on the fact
that three of the quark masses happen to be light, so that QCD
exhibits an approximate chiral symmetry. The Goldstone bosons
associated with the spontaneous breakdown of this symmetry represent the
essential degrees of freedom at low energies. I emphasize the universal
structure of the corresponding effective Lagrangian, which only depends on the
symmetry properties of the theory, and then discuss the physics of the
effective coupling constants. The implications of the effective theory
for the masses of the light quarks are analyzed in some detail.}

\section{Introduction}
The standard low energy analysis of
scattering amplitudes or current matrix
elements relies on the {\it Taylor series expansion} in powers
of the momenta.
The electromagnetic form factor of the pion, for instance, may be
exanded in powers of the momentum transfer $t$.
In this case, the first two Taylor coefficients are related to the total charge
of the particle and to the mean square radius of the charge distribution,
respectively,
\be\label{taylor}
f_{\pi^+}(t) = 1 + \frac{1}{6}\, \langle
r^2\rangle\rule[-0.3em]{0em}{1em}_{\!\pi^+} t + O(t^2)\fs \ee
Scattering lengths and effective ranges are analogous low energy
constants occurring in the Taylor series expansion of scattering amplitudes.

   For the straightforward expansion in powers of the momenta to hold it is
essential that the theory does not contain massless particles. The exchange of
photons, for example, gives rise to Coulomb scattering, described by an
amplitude of
the form $e^2/(p'-p)^2$ which does not admit a Taylor series expansion. Now,
QCD does not contain massless particles, but it does contain very light ones:
pions. The occurrence of light particles gives rise to singularities in the low
energy domain which limit the range of validity of the Taylor series
representation. The form factor $f_{\pi^+}(t)$ contains a cut starting
at $t=4 M_\pi^2$, such that the formula (\ref{taylor}) provides an adequate
representation only for $t\ll 4 M_\pi^2$. To extend this representation to
larger momenta, we need to account for the singularities generated by the
pions. This can be done, because the reason why $M_\pi$ is so small is
understood: The pions are the Goldstone bosons of a hidden, approximate
symmetry.\cite{Nambu}

The low energy singularities generated by the
remaining members
of the pseudoscalar octet $(K^\pm, K^0, \bar{K}^0, \eta)$ can be dealt with in
the same manner, exploiting the fact that the Hamiltonian of QCD is
approximately invariant under chiral $\mbox{SU(3)}_L\!\times\!\mbox{SU(3)}_R$.
If the three
light quark flavours $u,d,s,$ were massless, this symmetry would be an exact
one. In reality, chiral symmetry is broken by the quark mass term ocurring in
the QCD--Hamiltonian
\be
H_{\QCD} = H_0 + H_1 \co\;\;\;
H_1  =  \int d^3x \{m_u \bar{u}u + m_d \bar{d} d + m_s \bar{s}s \}\fs
\ee
For yet unknown reasons, the masses $m_u, m_d, m_s$ happen to be
small, so that
$H_1$ may be treated as a perturbation. First order perturbation theory shows
that the expansion of the square of the pion mass in powers of $m_u, m_d, m_s$
starts with
\be\label{e3}
M^2_{\pi^+} = (m_u + m_d) B +\ldots
\ee
while, for the kaon, the leading term involves the mass of the strange quark,
\be\label{e4}
M^2_{K^+}  =  (m_u + m_s) B + \ldots\co \;\;\;\;\;
M^2_{K^0}  =  (m_d + m_s) B + \ldots
\ee
This explains why the pseudoscalar octet contains the eight lightest hadrons
and why the mass pattern of this multiplet very strongly breaks eightfold way
symmetry: $M^2_\pi, M^2_K$ and $M^2_\eta$ are proportional to combinations of
quark masses, which are small but very different from one another, $m_s \gg m_d
> m_u$. For all other multiplets of SU(3), the main contribution to the mass is
given by the eigenvalue of $H_0$ and is of order $\Lambda_{\QCD}$,
while $H_1$ merely generates a correction which splits the
multiplet, the state with the largest matrix element of $\bar{s}s$ ending up at
the top.

   The effective field theory combines the expansion in powers of momenta with
the expansion in powers of $m_u, m_d, m_s$. The resulting new improved Taylor
series, which explicitly accounts for the singularities generated by the
Goldstone bosons, is referred to as chiral perturbation theory
(ChPT).\cite{ChPT}
It provides a solid mathematical basis for what used to be called the "PCAC
hypothesis".

It does not appear to be possible to account for the singularities generated by
the next heavier bound states, the vector mesons, in an equally satisfactory
manner. The mass of the $\rho$--meson is of the order of the scale of QCD and
cannot consistently be treated as a small quantity. Although the vector meson
dominance hypothesis does lead to valid estimates (an example is given below),
a coherent framework which treats these estimates as leading terms of a
systematic approximation scheme is not in sight.

\section{U(1)--anomaly and mass of the $\eta'$}
\label{U(1)}
I add a few remarks concerning the singlet axial current
\bdm A_\mu^0=\ubar\gamma_\mu\gamma_5 u+\dbar\gamma_\mu\gamma_5 d
+\sbar\gamma_\mu\gamma_5 s\fs\edm
 It is crucial
for QCD to be consistent with the observed low energy structure that the Ward
identities obeyed by this current contain an anomaly. At the level of
classical field theory, the Lagrangian is invariant under the full chiral
group $\mbox{U(3)}_L\!\times\!\mbox{U(3)}_R$ if the quark masses are turned off,
but this property does not give rise to a corresponding symmetry of the quantum
theory, because the measure entering
the path integral fails to be symmetric: The conservation law for
the singlet current contains an additional term,
\bdm \partial^\mu A_\mu^0=2\,m_u\,\ubar i\gamma_5 u+ 2\,m_d\,\dbar i\gamma_5 d+
2\,m_s\,\sbar i\gamma_5 s+
\frac{3}{8\pi^2}\,
\mbox{tr}\,G_{\mu\nu}\tilde{G}^{\mu\nu}\co\edm
where the matrix $G_{\mu\nu}$ is the gluon field strength.
Accordingly, there are only 8
rather
than 9 Goldstone bosons. The lightest pseudoscalar with the quantum numbers
of an SU(3)--singlet, the $\eta'$, remains massive if the quark masses are
turned off.

There is an old paradox in this connection, which arises as follows. Consider
the euclidean path integral representation for the two--point function
$\lvac T A_\mu(x) P(y)\rvac$, where
$P=\ubar i\gamma_5 u+\dbar i\gamma_5 d
+\sbar i\gamma_5 s$ is the pseudoscalar singlet operator.  The Ward
identity obeyed by this correlation function implies that, in the chiral
limit, the Fourier transform
contains a pole at zero momentum -- unless the contribution from the
anomaly term is different from zero. Now, at zero momentum, the
Fourier transform represents the integral over all of space, so that the
contribution
from the anomaly term is proportional to the winding number of the gluon field,
\bdm \nu =\frac{1}{16\pi^2}\int \!\!d^4\!x\,
\mbox{tr}\,G_{\mu\nu}\tilde{G}^{\mu\nu}\fs\edm
Hence the anomaly can prevent the occurrence of a massless
pseudoscalar SU(3)--singlet state only if the path integral receives
contributions from field configurations with nontrivial topology.
For such field configurations, however, the Dirac operator of a
massless quark possesses
zero modes. If the winding number of the gluon field is different from zero,
the Dirac determinant therefore tends
to zero when the quark masses are turned off, so that only topologically
trivial configurations can contribute to the path integral.
This seems to indicate that the anomaly term does not protect the $\eta'$ from
becoming massless in the chiral limit.

The puzzle is solved
in ref.~\cite{Leutwyler Smilga}, where it is demonstrated that the suppression
of nontrivial topologies only occurs at finite
volume and disappears if the limit $V\!\rightarrow\!\infty$ is taken before the
quark masses are sent to zero. The mechanism at work is very similar to the
one responsible for the fact that spontaneous
breakdown of an exact symmetry does not occur at finite volume.
I did discuss the matter
in some detail during the lectures, but omit this part
in the present notes, referring the reader to the paper quoted above.

In the large--$N_c$ limit, the quark--loop graph that gives rise to the
anomaly in the divergence of the singlet axial current is suppressed. The Ward
identity for the two--point function $\lvac T A_\mu P\rvac$ does then imply
that, in the chiral limit, a ninth Goldstone boson occurs: The mass of
the $\eta'$ disappears if $N_c$ is sent to infinity.\cite{Large Nc}
The implications for the structure of the effective theory are
very interesting and will briefly be discussed in section
\ref{theory} C.

\section{Effective low energy theory of QCD}
The effective low energy theory replaces the quark and gluon fields of QCD by a
set of pseudoscalar fields describing the degrees of freedom of the Goldstone
bosons $\pi, K, \eta$. It is convenient to collect these fields in a $3\times3$
matrix $U(x)\in \mbox{SU(3)}$. Accordingly, the Lagrangian of QCD is replaced
by an
effective Lagrangian which only involves the field U$(x)$ and its derivatives.
The most remarkable point here is that this procedure does not mutilate the
theory: If the effective Lagrangian is chosen properly, the effective theory is
mathematically equivalent to QCD.\cite{Weinberg Physica,GL SU(2)}

On the level of the effective Lagrangian, the combined expansion introduced
above amounts to an expansion in powers of derivatives and powers of the quark
mass matrix
\bdm
{\cal M} = \left(
\begin{array}{ccc}
 m_u &     &     \cr
     & m_d &     \cr
     &     &  m_d
\end{array}
\right)\fs
\edm
Lorentz invariance and chiral symmetry very strongly constrain the form of
the terms occurring in this expansion. Counting ${\cal M}$ like two powers of
momenta, the expansion starts at $O(p^2)$ and only contains even terms
\bdm
{\cal L}_\eff= {\cal L}_\eff^{(2)} +
{\cal L}_\eff^{(4)} + {\cal L}_\eff^{(6)} +
\ldots
\edm
The leading contribution is of the form
\be\label{eff Lag}
{\cal L}_\eff^{(2)} = \frac{F^2_\pi}{4} \mbox{tr} \{
\partial_\mu U^+ \partial^\mu U \} +
\frac{F^2_\pi B}{2} \mbox{tr} \{ {\cal M} (U+U^+) \}
\ee
and involves two independent coupling constants -- the pion decay constant
$F_\pi$ and the constant $B$ occurring in the mass formulae (\ref{e3}),
(\ref{e4}).
The expression (\ref{eff Lag}) represents a compact summary of the soft pion
theorems
established in the 1960's: The leading terms in the chiral expansion of the
scattering amplitudes and current matrix elements are given by the tree graphs
of this Lagrangian.

At order $p^4$, the effective Lagrangian contains terms with four derivatives
such as
\be
{\cal L}_\eff^{(4)} = L_1 [ \mbox{tr} \{ \partial_\mu U^+
\partial^\mu U \} ]^2 + \ldots
\ee
as well as terms with one or two powers of ${\cal M}$. Altogether, ten coupling
constants occur,\cite{GL SU(3)} denoted $L_1, \ldots, L_{10}$. Four of
these are needed
to specify the scattering matrix to first nonleading order. The terms of order
${\cal M}^2$ in the meson mass formulae (\ref{e3}), (\ref{e4}) involve another
three of
these constants. The remaining three couplings concern current matrix elements.

As an illustration, consider again the e.m. form factor $f_{\pi^+}(t)$. To
order $p^2$, the chiral representation reads~\cite{GL form factors}
\be\label{chiral representation}
f_{\pi^+} (t) =
1 + \frac{t}{F^2_\pi} \{2L_9 + 2 \phi_\pi(t) + \phi_K(t)\}
+ O(t^2, t {\cal M})\fs
\ee
In this example, the leading term (tree graph of ${\cal
L}_\eff^{(2)}$) is trivial. At order $p^2$, there are two
contributions: The term linear in $t$ arises from a tree graph of ${\cal
L}_\eff^{(4)}$ and involves the coupling constant $L_9$, while
the functions $\phi_\pi (t)$ and $\phi_K (t)$ originate in one--loop graphs
generated by ${\cal L}_\eff^{(2)}$. The loop integrals contain
a logarithmic divergence which is absorbed in a renormalization of $L_9$ -- the
net result for $f_{\pi^+} (t)$ is independent of the regularization used. The
representation (\ref{chiral representation}) shows how the straightforward
Taylor series (\ref{taylor}) is modified
by the singularites due to $\pi \pi$ and $K\bar{K}$ intermediate states. At the
order of the chiral expansion we are considering here, these singularities are
described by the one--loop integrals $\phi_\pi(t), \phi_K(t)$ which contain
cuts starting at $t=4 M_\pi^2$ and $t = 4 M^2_K$, respectively. The result
(\ref{chiral representation}) also shows that chiral symmetry does not determine the pion charge
radius: Its
magnitude depends on the value of the coupling constant $L_9$ -- the effective
Lagrangian is consistent with chiral symmetry for any value of the coupling
constants.
The symmetry,
however, {\it relates} different observables. In particular, the slope of the
$K_{l_3}$ form
factor $f_+(t)$ is also fixed by $L_9$. The experimental value of this
slope, $\lambda_+ = 0.030$, can therefore be used to first determine the
magnitude of $L_9$ and then to calculate the pion charge radius. This gives
$\langle r^2\rangle\rule[-0.3em]{0em}{1em}_{\!\pi^+} = 0.42$ fm$^2$, to be
compared with the experimental result, 0.44 fm$^2$.

In the case of the neutral kaon, the representation analogous to eq.\
(\ref{chiral representation}) reads \be
f_{K^0} (t) = \frac{t}{F^2_\pi} \{ - \phi_\pi (t) + \phi_K (t) \} + O(t^2, t
{\cal M})\fs
\ee
A term of order one does not occur here because the charge vanishes and there
is no contribution from ${\cal L}^{(4)}_\eff$, either. Chiral
perturbation theory thus provides a parameter free prediction in terms of the
one--loop integrals $\phi_\pi(t), \phi_K(t)$. In particular, the slope
of the form factor is given by~\cite{GL form factors}
\be
\langle r^2\rangle\rule[-0.3em]{0em}{1em}_{\!K^0} = - \frac{1}{16\pi^2F^2_\pi}
\ln \frac{M_K}{M_\pi} = - 0.04 \;\mbox{fm}^2
\ee
to be compared with the experimental value $- 0.054 \pm 0.026$ fm$^2$.\\

\section{Universality}
\label{uni}
The properties of the effective theory are governed by the hidden symmetry,
which is responsible for the occurrence of Goldstone bosons. In
particular, the form of the effective Lagrangian only depends on the symmetry
group G of the Hamiltonian and on the subgroup $\mbox{H}\subset \mbox{G}$,
under which the ground state is invariant. The Goldstone bosons live on the
difference between the two
groups, i.e., on the coset space G/H. The specific dynamical properties of the
underlying theory do not play any role. To discuss the consequences of this
observation,
I again assume that G is an exact symmetry.

In the case of QCD with two
massless quarks, $\mbox{G}\!=\!\mbox{SU(2)}_R\!\times\!\mbox{SU(2)}_L$ is
the group of
chiral isospin rotations, while $\mbox{H}\!=\!\mbox{SU(2)}$ is the ordinary
isospin
group.
The Higgs model is another example of a theory with spontaneously broken
symmetry. It plays a crucial role in the Standard Model, where it describes
the generation of mass. The model involves a scalar field
$\vec{\phi}$ with four components. The Hamiltonian is invariant under
rotations of the vector $\vec{\phi}$, which form the group $\mbox{G}\! =\!
\mbox{O(4)}$.
Since the
field picks up a vacuum expectation value, the
symmetry is spontaneously broken to the subgroup of those rotations that
leave the vector $\lvac\vec{\phi}\rvac$ alone, $\mbox{H} \!=\!\mbox{O(3)}$.
It so happens that these groups are the same as those above,
relevant for QCD.\footnote{The structure of the effective Lagrangian
rigorously follows from the
Ward identities for the Green functions of the currents, which also reveal the
occurrence of anomalies.\cite{found} The
form of the Ward identities is controlled by
the structure of G and H in the infinitesimal neighbourhood of the
neutral element. In this sense, the symmetry groups of the two models are the
same: O(4) and O(3) are locally
isomorphic to SU(2)$\times$SU(2) and
SU(2), respectively. }
The fact that the symmetries are the same implies that
the effective field theories are identical: (i) In either
case, there are three
Goldstone bosons, described by a matrix field $U(x)\in\mbox{SU(2)}$. (ii) The
form of the effective Lagrangian is precisely the same.
In particular, the expression
\bdm
{\cal L}_{\eff}^{(2)} = \mbox{$\frac{1}{4}$}F_\pi^2 \mbox{tr} \{
\partial_\mu U^+ \partial^\mu U \}
\edm
is valid in either case. At the level of the effective theory, the
only
difference between these two physically quite distinct models is that
the numerical values of the effective coupling constants are different.
In the case of QCD, the one occurring at leading order of the
derivative expansion is the pion decay constant, $F_\pi\simeq
93\,\mbox{MeV}$, while in the Higgs model, this coupling constant is larger
by more than three orders of magnitude, $F_\pi\simeq
250\;\mbox{GeV}$. At next--to--leading order, the effective coupling constants
are also different; in particular, in QCD, the anomaly coefficient is equal to
$\mbox{N}_c$, while in the Higgs model, it vanishes.

As an illustration, I compare the condensates of the two theories, which
play a role
analogous to the spontaneous magnetization $\langle\Mvec\rangle$ of a
ferromagnet (or the staggered magnetization of an antiferromagnet).
At low temperatures, the magnetization singles out a direction -- the ground
state spontaneously breaks the symmetry
of the Hamiltonian with respect to rotations. As the system is heated, the
spontaneous magnetization decreases, because the thermal disorder acts against
the alignment of the spins. If the temperature is high enough, disorder
wins, the spontaneous magnetization disappears and rotational symmetry is
restored. The temperature at which this happens is the Curie temperature.
Quantities, which allow one to distinguish the ordered from the disordered
phase are called {\it order parameters}. The magnetization is the prototype of
such a parameter.

In QCD, the most important order parameter (the one of lowest dimension) is the
quark condensate. At nonzero temperatures, the condensate is given
by the thermal expectation
value \bdm \langle\ubar u\rangle_{\hspace{-0.05em}\mbox{\raisebox{-0.2em}
{\scriptsize $T$}}} =\frac{\mbox{Tr}\{\,\ubar u
\exp (-\,H/kT)\}}{ \mbox{Tr}\{\exp(-\,H/kT)\} }\fs
\edm
The condensate melts if the temperature is
increased. At a critical temperature, somewhere in the range
$140\,\mbox{MeV}\!<\!T_c\!<\!\mbox{180}\;\mbox{MeV}$, the quark condensate
disappears and chiral symmetry is restored. The same qualitative
behaviour also occurs in the Higgs model, where the expectation value
$\langle\,\vec{\phi}\,
\rangle_{\hspace{-0.05em}\mbox{\raisebox{-0.2em}
{\scriptsize $T$}}}$ of the scalar field represents the most prominent order
parameter.

At low temperatures, the thermal trace is dominated by
states of low energy. Massless particles generate contributions which are
proportional to powers of the temperature, while massive ones like the
$\rho$--meson are suppressed by the corresponding Boltzmann factor,
$\exp(-M_\rho/kT)$. In the case of a spontaneously broken symmetry,
the massless particles are the Goldstone bosons and their contributions may be
worked out by means of effective field theory. For the quark condensate, the
calculation has been done,\cite{Gerber} up to and including terms of order
$T^6$: \bdm\langle\ubar
u\rangle_{\hspace{-0.05em}\mbox{\raisebox{-0.1em} {\scriptsize $T$}}} =
\lvac\ubar u\rvac\!
\left\{1\,-\,\frac{T^2}
{8F_\pi^2 }
\,-\,\frac{T^4}{384F_\pi^4}
\,-\,\frac{T^6}{288F_\pi^6}
\, \ln(T_1/T)
\,+\,O(T^8)\right\}\fs\edm
The formula is exact -- for massless quarks, the temperature scale relevant
at low $T$ is the pion decay constant. The additional logarithmic scale $T_1$
occurring at order $T^6$ is determined by the effective coupling constants that
enter the expression for the effective Lagrangian
of order $p^4$. Since these are known from the phenomenology of $\pi\pi$
scattering, the value of $T_1$ is also known:
$T_1=470\pm110\;\mbox{MeV}$.

Now comes the point I wish to make. The effective Lagrangians
relevant for QCD and for the Higgs model are the same. Since the
operators of which we are considering the expectation values also transform in
the
same manner, their low temperature expansions are identical. The above formula
thus holds, without any change whatsoever, also for the Higgs condensate,
\bdm \langle\,\vec{\phi}\,\rangle_{\hspace{-0.1em}\mbox{\raisebox{-0.2em}
{\scriptsize $T$}}} =
\lvac\vec{\phi}\rvac\!
\left\{1\,-\,\frac{T^2}
{8F_\pi^2 }
\,-\,\frac{T^4}{384F_\pi^4}
\,-\,\frac{T^6}{288F_\pi^6}
\, \ln(T_1/T)
\,+\,O(T^8)\right\}\fs\edm
In fact, the universal term of order $T^2$ was discovered in the framework of
this model, in connection with work on the electroweak phase
transition.\cite{Binetruy}

These examples illustrate the physical nature of effective theories: At long
wavelength, the microscopic structure does not play any role. The behaviour
only depends on those degrees of freedom that require little
excitation
energy. The hidden symmetry, which is responsible for the absence of an
energy gap and for the occurrence of Goldstone bosons, at the same time also
determines their low energy properties. For this reason, the form of
the effective Lagrangian is controlled
by the symmetries of the system and is, therefore, universal.
The microscopic structure of the underlying theory exclusively manifests itself
in the numerical values of the effective coupling constants.
The temperature expansion also clearly exhibits the limitations of
the method. The truncated series can be trusted only at low temperatures,
where the first term represents the dominant contribution. According to the
above formula, the quark condensate drops to about half of the vacuum
expectation value when the temperature reaches
$160\;\mbox{MeV}$ -- the formula does not make much sense beyond this
point. In particular, the behaviour of
the quark condensate in the vicinity of the chiral phase transition is
beyond the reach of the effective theory discussed here.

\section{Physics of the effective coupling constants}
\label{pecc}
One of the main problems encountered in the effective Lagrangian
approach is the occurrence of an entire fauna of effective coupling constants.
If these constants are treated as totally arbitrary parameters, the predictive
power of the method is equal to zero -- as a bare minimum, an estimate of their
order of magnitude is needed.

{\em Chiral scale.} Let me first drop the masses of the light quarks and send
the heavy ones
to infinity. In this limit, QCD is a theoretician's paradise: A theory without
adjustable dimensionless parameters. In particular, the effective coupling
constants $F_\pi, B, L_1, L_2, \ldots$ are given by pure numbers multiplying
powers of $\Lambda_{\QCD}$. In principle, the numbers are
calculable -- the available, admittedly crude evaluations of $F_\pi$ and $B$ on
the lattice demonstrate that the calculation is even feasible.

As discussed
above, the coupling constants $L_1, \ldots, L_{10}$ are renormalized by the
logarithmic divergences occurring in the one--loop graphs. This property
sheds considerable light on the structure of the chiral expansion and provides
a rough estimate for the order of magnitude of the effective coupling
constants.\cite{Georgi Manohar} The point is
that the contributions generated by the loop
graphs are smaller than the leading (tree graph) contribution only for momenta
in the range $|\hspace{0.05em}p\hspace{0.1em}|\lsim  \Lambda_\chi$, where
\be\label{chiral scale}
\Lambda_\chi \equiv 4 \pi F_\pi/ \sqrt{N_f}
\ee
is the scale occurring in the coefficient of the logarithmic divergence ($N_f$
is the number of light quark flavours). This indicates that the derivative
expansion is an expansion in powers of $(p/\Lambda_\chi)^2$ with coefficients
of order one. The stability argument also applies to the expansion in powers of
$m_u, m_d$ and $m_s$, indicating that the relevant expansion parameter is given
by $(M_\pi/ \Lambda_\chi)^2$ and $(M_K/\Lambda_\chi)^2$, respectively.

{\em Resonances.} A more quantitative picture can be obtained along the
following lines. Consider
again the e.m. form factor of the pion and compare the chiral representation
(\ref{chiral representation}) with the dispersion relation
\be
f_{\pi^+}(t) = \frac{1}{\pi} \int^{\infty}_{4M^2_\pi} \frac{dt'}{t'-t}
\mbox{Im} f_{\pi^+} (t') \fs
\ee
In this relation, the contributions $\phi_\pi, \phi_K$ from the
one--loop graphs of ChPT correspond to $\pi \pi$ and $K \bar{K}$ intermediate states. To
leading order in the chiral expansion, the corresponding imaginary parts are
slowly rising functions of $t$. The most prominent contribution on the r.h.s.,
however, stems from the region of the $\rho$--resonance which nearly saturates
the integral: The vector meson dominance formula, $f_{\pi^+} (t) =
(1-t/M_\rho^2)^{-1}$, which results if all other contributions are dropped,
provides a perfectly decent representation of the form factor for small values
of $t$. In particular, this formula predicts $\langle r^2\rangle_{\pi^+} =
0.39$ fm$^2$,
in satisfactory agreement with observation (0.44 fm$^2$). This implies that the
effective coupling constant $L_9$ is approximately given by~\cite{GL SU(2)}
\be
L_9 = \frac{F^2_\pi}{2M^2_\rho}\fs
\ee
In the channel under consideration, the pole due to $\rho$ exchange thus
represents the dominating low energy singularity -- the $\pi \pi$ and $K
\bar{K}$ cuts merely generate a small correction. More generally, the validity
of the vector meson dominance formula shows that, for the e.m. form factor,
the scale of the derivative expansion is set by $M_\rho = 770$ MeV.

Analogous estimates can be given for all effective coupling constants at order
$p^4$, saturating suitable dispersion relations with contributions from
reso\-nan\-ces,\cite{Ecker Gasser Pich de Rafael} for instance:
\be\label{e16}
L_5 = \frac{F_\pi^2}{4M^2_S}\co\hspace{1cm}
 L_7 = - \frac{F^2_\pi}{48M^2_{\eta'}}\co
\ee
where $M_S \simeq 980$ MeV and $M_{\eta'} = 958$ MeV are the masses of the
scalar octet and pseudoscalar singlet, respectively. In all those cases where
direct phenomenological information is available, these estimates do
remarkably well. I conclude that the observed low energy structure is dominated
by the poles and cuts generated by the lightest particles -- hardly a surprise.
In some channels, the scale of the chiral expansion is set by $M_\rho$, in
others by the masses of the scalar or pseudoscalar resonances occurring around
1 GeV. This confirms the rough estimate (\ref{chiral scale}). The cuts
generated
by Goldstone pairs are significant in some cases and are negligible in others,
depending on the numerical value of the relevant Clebsch--Gordan coefficient.
If this coefficient turns out to be large, the coupling constant in question is
sensitive to the renormalization scale used in the loop graphs. The
corresponding pole dominance formula is then somewhat fuzzy, because the
prediction depends on how the resonance is split from the continuum underneath
it.

The above quantitative estimates of the scale of the expansion
explain why
it is justified to treat $m_s$ as a perturbation.\cite{Leutwyler1990} At order
$p^4$, the symmetry
breaking part of the effective Lagrangian is determined by the coupling
constants $L_4, \ldots, L_8$. These constants are immune to the low energy
singularities generated by spin 1 resonances, but are affected by the exchange
of scalar or pseudoscalar particles. Their magnitude is therefore determined
by the scale $M_S \simeq M_{\eta'} \simeq 1$ GeV [see eq.\ (\ref{e16})].
Accordingly,
the expansion in powers of $m_s$ is controlled by the parameter $(M_K/M_S)^2
\simeq \frac{1}{4}$. Disregarding the contributions generated by the
one--loop graphs, the asymmetry in the decay constants, for example,
is determined
by $L_5$:
\be\label{DeltaF}\frac{F_K}{F_\pi}=1+\frac{4(M_K^2-M_\pi^2)}{F_\pi^2}\,L_5
+\mbox{chiral logs}\fs\ee
The term "chiral logs" stands for the logarithms characteristic of
chiral perturbation theory. In the present case they arise from the
two--particle continuum underneath the resonance. Retaining only the
resonance contribution, we obtain
\be
\frac{F_K - F_\pi}{F_K} = \frac{M^2_K - M^2_\pi}{M^2_S}+\ldots
\ee
This shows that the breaking of the chiral and eightfold way symmetries is
controlled by the mass ratio of the Goldstone bosons to the non--Goldstone
states of spin zero -- in ChPT, the observation that the Goldstones are the
lightest hadrons thus acquires quantitative significance.

\section{Mass pattern of the light quarks}
In the remainder of these lectures, I concentrate on one particular application
of chiral perturbation theory and discuss the implications for the masses of
the light quarks. The lowest order mass formulae for the
Goldstone bosons, eqs.\ (\ref{e3}) and (\ref{e4}), imply that the quark mass
ratios are approximately given by
\bdm
\frac{m_u}{m_d}\simeq
\frac{M_{\pi^+}^2-M_{K^0}^2+M_{K^+}^2}{M_{\pi^+}^2+M_{K^0}^2-M_{K^+}^2}
\simeq 0.66\co\;\;\;\;
\frac{m_s}{m_d}\simeq
\frac{M_{K^0}^2+M_{K^+}^2-M_{\pi^+}^2}{M_{K^0}^2-M_{K^+}^2+M_{\pi^+}^2}
\simeq 20\co\edm
to be compared with $m_\mu/m_e\simeq 200$.
These numbers represent rough estimates. The corrections generated by the
higher order terms in the mass
formulae as well as those due to the electromagnetic interaction are treated
below -- they do not significantly modify the above ratios.

First,
however, I wish to discuss some qualitative aspects of this pattern.
For this purpose, I need a crude estimate for the absolute magnitude of the
light quark masses, which may be obtained with the following simple argument.
The mass differences between $m_u$, $m_d$ and $m_s$ are responsible for the
splittings observed within the multiplets of SU(3). The observed multiplet
pattern shows that, replacing a $u$-- or a $d$--quark by an
$s$--quark, the mass of the bound state increases by about 100 or 200 MeV.
Applying the rule of thumb, we infer that the mass differences $m_s-m_u$ and
$m_s-m_d$ are of this order of magnitude. Since $m_u$ and
$m_d$ are small compared to $m_s$, the mass of the strange
quark must be of this order, say $m_s\simeq 150$ MeV. With the above ratios,
this gives $m_d\simeq
\frac{1}{20}\hspace{0.1em}m_s\simeq 7.5$ MeV and $m_u\simeq \frac{2}{3}
\hspace{0.05em}m_d\simeq 5$ MeV. I
emphasize
that these estimates only concern the order of magnitude and I will discuss
our present knowledge at a quantitative level later on.

The first conclusion to draw is that $m_u$ and $m_d$ are surprisingly small.
In particular, the mass of the proton is large compared to the sum of the
masses of the quarks it consists of. Indeed, the mass of the proton does not
tend to zero for $m_u,m_d\!\rightarrow\!0$. The amount by which it decreases
is known from $\pi N$--scattering: $\sigma=\langle N|\,m_u\ubar u+m_d\dbar d\,|
N\rangle= 45\pm 8$ MeV. This shows that the masses occurring in the Lagrangian
of QCD are quite different from those used in the various bound state models,
$m_{constituent}\simeq\frac{1}{3}M_p\simeq 300$ MeV.

An equally striking aspect of the above pattern is
that the three masses are very different. In particular, the value for
$m_u/m_d$ shows that the masses of the $u$-- and $d$--quarks are quite
different. This appears to be in conflict with the oldest and best
established internal symmetry of particle physics, isospin. Since $u$ and $d$
form an $I\!=\!\frac{1}{2}$ multiplet, isospin is a symmetry of the QCD
Hamiltonian only if $m_u\!=\!m_d$.

The resolution of the paradox
is that $m_u,m_d$ are very small. Disregarding the e.m.\ interaction, the
strength of isospin breaking is determined by the magnitude of $|m_u\!-\!m_d|$,
not by the relative size $m_u/m_d$.
The fact that $m_d$ is larger than $m_u$ by a few MeV implies, for instance,
that the neutron is heavier than the proton by a few MeV. Compared with the
mass of the proton, this amounts to a fraction of a per cent.
In the case
of the kaons, the relative mass splitting $(M_{K^0}^2\!-\!M_{K^+}^2)/M_{K^+}^2$
is more important, because the denominator is smaller here: The effect is of
order $(m_d-m_u)/(m_u+m_s)\simeq 0.02$, but this is still a small number. One
might
think that for the pions, where the square of the mass is proportional to
$m_u+m_d$, the relative mass splitting should be large, of order
$(M_{\pi^0}^2\!-\!M_{\pi^+}^2)/M_{\pi^+}^2\propto
(m_d\!-\!m_u)/(m_d\!+\!m_u)\simeq 0.3 $, in flat contradiction with
observation. It so happens, however, that the pion matrix elements of
the isospin breaking part of the Hamiltonian,
$\frac{1}{2}(m_u\!-\!m_d)(\ubar u-\dbar d)$, vanish
because the group
SU(2) does not have a $d$--symbol. This implies that the mass difference
between
$\pi^0$ and $\pi^+$ is of second order in $m_d\!-\!m_u$ and therefore tiny. The
observed
mass difference is almost exclusively due to the electromagnetic self energy of
the $\pi^+$. So, the above quark mass pattern is perfectly consistent with
the fact that isospin is an almost exact symmetry of the strong
interaction: The matrix elements of the term $\frac{1}{2}(m_u\!-\!m_d)(\ubar
u-\dbar d)$ are very small compared with those of $H_0$. In particular, the
pions are protected from isospin breaking.

QCD also explains another puzzle: Apparently, the mass splittings in the
pseudoscalar octet are in conflict with the claim that SU(3) represents a
decent approximate symmetry. This seems to require $M_{K}^2\!\simeq\!
M_{\pi}^2$,
while experimentally, $M_{K}^2\!\simeq\! 13 M_{\pi}^2$. The first
order mass
formulae yield $M_{K}^2\!/\!M_{\pi}^2\!=\!(m_s\hspace{-0.1em}+
\hspace{-0.02em}\hat{m})
\hspace{-0.07em}/\hspace{-0.07em}(m_u\hspace{-0.1em}+\hspace{-0.02em}m_d)$,
where
$\hat{m}\!=\!\frac{1}{2}(m_u\hspace{-0.1em}+\hspace{-0.02em}m_d)$ is the mean
mass of $u$ and $d$. The
kaons are much heavier than the pions, because it so happens
that $m_s$ is much larger than $\hat{m}$. For SU(3) to be a
decent approximate symmetry, it is not necessary that the difference
$m_s-\hat{m}$ is small with respect to the sum $m_s+\hat{m}$, because the
latter does
not represent the relevant mass scale to compare the symmetry breaking with.
If the quark masses were of the same order of magnitude as the electron
mass, SU(3) would be an essentially perfect symmetry of QCD; even in that
world $m_s\!\gg\! \hat{m}$ implies that the ratio $M_K^2/M_\pi^2$ strongly
differs from $1$.
The strength of SU(3) breaking does not manifest itself in the mass
ratios of the pseudoscalars, but in the symmetry relations between the matrix
elements
of the operators $\ubar u,\,\dbar d,\,\sbar s$, which are used in the
derivation of the above mass formulae. The asymmetries in these are analogous
to the one seen in the matrix
elements of the axial vector currents,
$F_K/F_\pi\!=\!1.22$, which represents an
SU(3) breaking of typical size. The deviation from the lowest order mass
formula,
\be
\label{i5}\frac{M_K^2}{M_\pi^2}=
\frac{m_s+\hat{m}}{m_u+m_d}\{1+\Delta_M\}\co\ee
is expected to be of the same order of magnitude,
$1\!+\!\Delta_M\!\leftrightarrow \!F_K/F_\pi$.

The Gell--Mann--Okubo formula yields a good check. The lowest order mass
formula for the $\eta$ reads
\be\label{i6} M_\eta^2=\mbox{$\frac{1}{3}$}(m_u+m_d+4m_s)B+\ldots\co\ee
so that the mass
relations for $\pi,K,\eta$ lead to $3M_\eta^2+M_\pi^2-4M_K^2\!=\!0$. The
accuracy within which this consequence of SU(3) symmetry holds is best seen
by working out the quark mass ratio $m_s/\hat{m}$ in two independent ways:
While the mass formulae for $K$ and $\pi$ imply
$m_s/\hat{m}\!=\!(2M_K^2\!-\!M_\pi^2)/M_\pi^2\!=\!25.9$, those for $\eta$
and $\pi$ yield $m_s/\hat{m}\!=\!\frac{1}{2}(3M_\eta^2\!-\!M_\pi^2)/M_\pi^2
\!=\!24.2$. These numbers are nearly the same --
the mass pattern of the pseudoscalar octet is
a showcase for the claim that SU(3) represents a decent
approximate symmetry of QCD, despite $M_K^2/M_\pi^2\simeq 13$.

\section{Mass formulae to second order}
Chiral perturbation theory allows us to calculate the Goldstone boson
masses to second order in $m_u,m_d,m_s$. The correction $\Delta_M$
occurring in eq.\ (\ref{i5}) is determined by the two coupling constants $L_5$
and $L_8$:~\cite{GL SU(3)} \be
\label{DeltaM}\Delta_M=
\frac{8(M_K^2-M_\pi^2)}{F_\pi^2}\,(2L_8-L_5)+\mbox{chiral logs}\fs\ee
The comparison with eq.\ (\ref{DeltaF}) confirms that the
symmetry
breaking effects in the decay constants and in the mass spectrum are of
similar nature. The calculation also reveals that the first order SU(3)
correction in the
mass ratio $(M_{K^0}^2-M_{K^+}^2)/(M_K^2-M_\pi^2)$ is the same as the one in
$M_K^2/M_\pi^2$:~\cite{GL SU(3)}
\be \frac{M_{K^0}^2-M_{K^+}^2}{M_K^2-M_\pi^2}=\frac{m_d-m_u}{m_s-\hat{m}}
\{1+\Delta_M +O(m^2)\}\fs\ee
In the double ratio
\be\label{defQ}
Q^2 \equiv \frac{M^2_K}{M_\pi^2}\cdot \frac{ M^2_K - M^2_\pi}{M^2_{K^0} -
M^2_{K^+}}\co \end{equation}
the first order corrections thus drop out, so that
the observed values of the meson masses provide a tight constraint on one
particular ratio of quark masses:
\be
Q^2 = \frac{m^2_s - \hat{m}^2}{m^2_d - m^2_u} \{ 1 + O (m^2) \}\fs
\end{equation}
The constraint may be visualized by
plotting the ratio
$m_s/m_d$ versus $m_u/m_d$.\cite{Kaplan Manohar}
Dropping
the higher order contributions, the
resulting curve takes the form of an ellipse:
\be\label{ellipse}
\left ( \frac{m_u}{m_d} \right)^2 + \,\frac{1}{Q^2} \left ( \frac{m_s}{m_d}
\right)^2 = 1\co
\end{equation}
with $Q$ as major semi--axis (the term $\hat{m}^2/m_s^2$ has been discarded,
as it is
numerically very small).

\section{Value of $Q$}
The meson masses occurring in the double ratio (\ref{defQ}) refer to pure QCD.
The Dashen theorem states that in the chiral limit, the electromagnetic
contributions to $M_{K^+}^2$ and $M_{\pi^+}^2$ are the same, while the self
energies of $K^0$ and $\pi^0$ vanish.
Since the contribution to the
mass difference between $\pi^0$ and $\pi^+$ from $m_d\!-\!m_u$ is
negligibly small, the masses in pure QCD are approximately given by
\bea (M_{\pi^+}^2)^{\QCD}\al\simeq\al
(M_{\pi^0}^2)^{\QCD}\simeq M_{\pi^0}^2\co\no (M_{K^+}^2)^{\QCD}\al\simeq\al
M_{K^+}^2-M_{\pi^+}^2+M_{\pi^0}^2\co\;\;\;(M_{K^0}^2)^{\QCD}\simeq M_{K^0}^2\co
\nonumber\eea
where $M_{\pi^0},M_{\pi^+},M_{K^0},M_{K^+}$ are the
observed masses. Correcting
for the electromagnetic self energies in this way, the lowest order formulae
become~\cite{Weinberg1977} \bea \label{i3}
\frac{m_u}{m_d} \al \simeq\al \frac{M^2_{K^+}-  M^2_{K^0} + 2 M^2_{\pi^0} -
M^2_{\pi^+}} {M^2_{K^0} - M^2_{K^+} + M^2_{\pi^+}}=0.55\co \\
\frac{m_s}{m_d} \al \simeq \al \frac{M^2_{K^0} + M^2_{K^+} - M^2_{\pi^+}}
{M^2_{K^0} - M^2_{K^+} + M^2_{\pi^+}}=20.1\fs\nonumber
\eea
The corresponding expression for the semi--axis $Q$ reads \be\label{QD}
\QD^{\;2}= \frac{(M_{K^0}^2+M_{K^+}^2-M_{\pi^+}^2+M_{\pi^0}^2)
(M_{K^0}^2+M_{K^+}^2-M_{\pi^+}^2-M_{\pi^0}^2)}
{4\,M_{\pi^0}^2\,(M_{K^0}^2-M_{K^+}^2+M_{\pi^+}^2-M_{\pi^0}^2)}\fs\ee
Numerically, this yields $\QD=24.2$. The corresponding ellipse is shown in
fig.\ 1 as a dash--dotted line.
\begin{figure}[t]
\centering
\mbox{\epsfysize=6.5cm \epsfbox{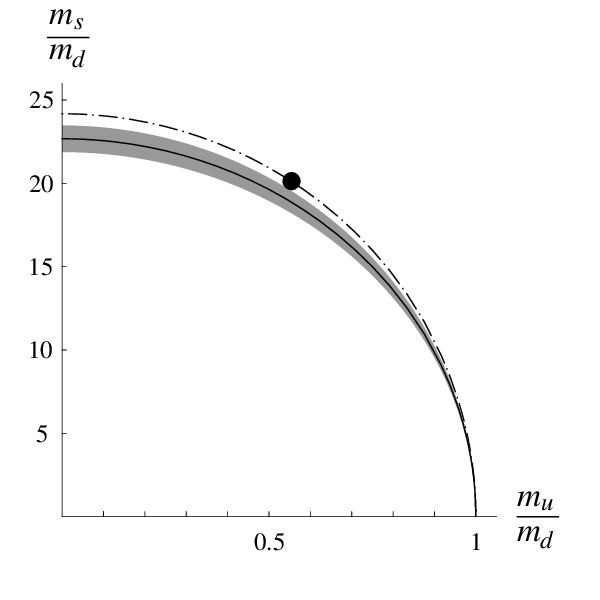} }
%\mbox{ \epsfbox{fitep1.eps} }
\parbox{11.9cm}{Figure 1:
Elliptic constraint. The dot indicates Weinberg's mass ratios.
The dash--dotted line represents the ellipse for the value $Q=24.2$ of the
semi--axis, obtained from
the mass
difference $K^0-K^+$ with the Dashen theorem. The full line and
the shaded region correspond to $Q=22.7\pm 0.8$, as required by
the observed rate of the decay $\eta\!\rightarrow\!\pi^+\pi^-\pi^0$.}
\end{figure}
For this value of the semi--axis, the curve passes through the point specified
by Weinberg's mass ratios, eq.\ (\ref{i3}).

The Dashen theorem is subject to corrections from higher order terms in the
chiral expansion. As usual, there are two categories of contributions: loop
graphs of order $e^2m$ and terms of the same order from the derivative
expansion
of the effective e.m.\ Lagrangian.
The Clebsch--Gordan coefficients occurring in
the loop graphs are known to be large, indicating that two--particle
intermediate
states generate sizeable corrections; the
corresponding chiral logarithms
tend to increase the e.m.\ contribution to the kaon mass
difference.\cite{Langacker}
The numerical result depends on the scale used when evaluating the logarithms.
In fact, taken by themselves, chiral logs are unsafe at any scale -- one at
the same time also needs to consider the contributions
from the terms of order $e^2m$
occurring in the effective Lagrangian. This is done in several
recent papers,\cite{DHW em self energies}$^-$\cite{Bijnens}
but the results are controversial.
The authors of ref.~\cite{DHW em self energies} estimate the contributions
arising from vector meson exchange and conclude
that these give rise to large corrections, increasing the
value $(M_{K^+}\!-\!M_{K^0})_{e.m.} \!=\! 1.3$ MeV predicted by Dashen to $2.3$
MeV. According to ref.~\cite{Urech}, however, the model used is in conflict
with chiral symmetry:
Although the perturbations due to vector meson exchange are enhanced
by a relatively small energy denominator, chiral symmetry prevents
them from being large. In view of this, it is puzzling that an
evaluation based on the ENJL model
yields an even larger effect,
$(M_{K^+}\!-\!M_{K^0})_{e.m.}\! \simeq\! 2.6\,\mbox{MeV}$.\cite{Bijnens}

Recently,
the electromagnetic self energies have been analyzed within
lattice QCD.\cite{Duncan Eichten Thacker} The
result, $(M_{K^+}\!-\!M_{K^0})_{e.m.} \!=\! 1.9$ MeV,
indicates that the corrections
to the Dashen theorem are indeed substantial, although not quite as large as
found in refs.~\cite{DHW em self energies,Bijnens}.
The uncertainties of the lattice result are of the same type as those
occuring in direct determinations of the quark masses with this method. The
mass difference between $K^+$ and
$K^0$, however, is predominantly due to $m_d\!>\!m_u$, not to the e.m.\
interaction. An error in the self energy of 20\% only affects the value of
$Q$ by about 3\%. The terms neglected when evaluating $Q^2$ with
the meson masses are of order $(M_K^2-M_\pi^2)^2/M_0^4$, where $M_0$ is
the mass scale relevant for the exchange of scalar or pseudoscalar
states, $M_0\!\simeq\! M_S\!\simeq\! M_{\eta'}$. The
corresponding error in the result for $Q$ is also of the order of 3\% --
the uncertainties in
the value $Q\!=\!22.8$ that follows from the lattice result are significantly
smaller than those obtained for the quark masses with the same method.
The implications of the above estimates for the value of
$Q$ are illustrated on the r.h.s. of fig.\ 2.
\begin{figure}[t] \centering
\mbox{\epsfysize=6.5cm \epsfbox{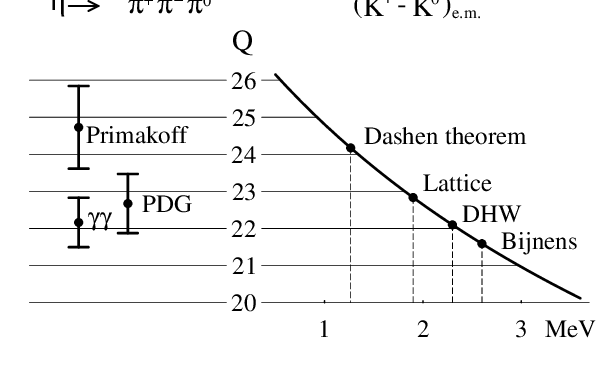} }
\parbox{11.9cm}{Figure 2:
The l.h.s. indicates the values of $Q$ corresponding to the various
experimental results for the rate
of the decay $\eta\!\rightarrow\!\pi^+\pi^-\pi^0$. The r.h.s. shows
the results for $Q$ obtained with three different theoretical estimates for
the electromagnetic self energy of the kaons.}\end{figure}

The isospin violating decay $\eta \rightarrow 3\pi$ allows an entirely
independent measurement of the semi-axis.\cite{GL eta} The transition
amplitude is much less sensitive to the
uncertainties associated with the electromagnetic interaction than the
$K^0\!-\!K^+$ mass difference: The e.m.\ contribution is
suppressed by chiral symmetry and is negligibly small.\cite{Sutherland} The
transition amplitude thus represents a sensitive probe of the
symmetry breaking generated by $m_d-m_u$. To lowest order in
the chiral expansion (current algebra), the amplitude of the transition
$\eta\!\rightarrow\!\pi^+\pi^-\pi^0$ is given by
\bdm A=-\frac{\sqrt{3}}{4}\,\frac{m_d-m_u}{m_s-\hat{m}}\,
\frac{1}{F_\pi^2}\,(s-\mbox{$\frac{4}{3}$}M_\pi^2)\co\edm
where $s$ is the square of the centre--of--mass energy of the charged pion
pair.
The corrections of first non--leading order (chiral perturbation
theory to one loop) are also known.
It is convenient to write the decay rate in the form
$\Gamma_{\eta \rightarrow \pi^+\pi^-\pi^0} \!=\!
\Gamma_0\,(\QD/Q)^4$, where $\QD$ is specified in eq.\ (\ref{QD}). As shown in
ref.~\cite{GL eta},
the one--loop calculation yields a parameter free prediction for the constant
$\Gamma_0$.
Updating the value of $F_\pi$, the numerical result reads
$\Gamma_0\!=\!168\pm50\,\mbox{eV}$.
Although the calculation includes
all corrections of
first non--leading order, the error bar is large. The problem originates in the
final state interaction, which strongly amplifies the transition probability
in part of the Dalitz plot. The one--loop calculation does account for this
phenomenon, but only to leading order in the low energy expansion.
The final state interaction is analyzed more accurately in two recent
papers,\cite{Kambor Wiesendanger Wyler,AL} which exploit the fact that
analyticity and unitarity determine the
amplitude up to a few subtraction constants. For these, the corrections to the
current algebra predictions are small, because they are barely
affected by the final state interaction. Although the dispersive framework
used in the two papers differs, the results are nearly the same: While
Kambor, Wiesendanger and Wyler obtain
$\Gamma_0\!=\!209\pm20\,\mbox{eV}$,
we get $\Gamma_0\!=\!219\pm22\,\mbox{eV}$.
This shows that the theoretical uncertainties of the dispersive calculation are
small. Since the decay rate is proportional to $Q^{-4}$, the transition
$\eta\!\rightarrow\!3\pi$ represents an
extremely sensitive probe, allowing a determination of $Q$ to an accuracy
of about
$2\frac{1}{2}\%$.

Unfortunately, however, the experimental situation is not
clear.\cite{PDG} The value of $\Gamma_{\eta\rightarrow
\pi^+\pi^-\pi^0}$ relies on the rate of the decay into two photons.
The two different methods of measuring $\Gamma_{\eta\rightarrow
\gamma\gamma}$ -- photon--photon collisions and
Primakoff effect -- yield conflicting results.
While the data based
on the Primakoff effect are in perfect agreement with the
number $Q= 24.2$ which follows from the Dashen theorem,
the $\gamma\gamma$ data
yield a significantly lower result (see l.h.s.\ of fig.\ 2).
The statistics is
dominated by the $\gamma\gamma$ data. Using the overall fit of the Particle
Data Group,
$\Gamma_{\eta\rightarrow\pi^+\pi^-\pi^0}\!=\!283\pm28\,\mbox{eV}$~\cite{PDG}
and adding errors quadratically, we obtain $Q\!=\!22.7\pm 0.8$, to be compared
with the value $Q=22.4\pm0.9$ given in ref.~\cite{Kambor Wiesendanger Wyler}.
The result appears to confirm the lattice
calculation.\cite{Duncan Eichten Thacker}
The above discussion
makes it clear that an improvement of the experimental situation concerning
$\Gamma_{\eta\rightarrow\gamma\gamma}$ is of considerable interest.

\section{A phenomenological ambiguity}\label{KM}
Chiral perturbation theory thus fixes one of the two quark mass ratios in
terms of the other, to within small uncertainties. The ratios themselves,
i.e. the position on the ellipse, are a more subtle issue. Kaplan and
Manohar~\cite{Kaplan Manohar} pointed out that the
corrections to the lowest order result, eq.\ (\ref{i3}),
cannot be determined on purely phenomenological grounds.
They
argued that these corrections might be large and that the $u$--quark might
actually be massless. This possibility is widely discussed in the
literature,\cite{Banks Nir Seiberg} because
the strong CP problem would then disappear.

The reason why phenomenology alone does not allow us to determine the
two individual ratios beyond leading order is the following.
The matrix \bdm m' = \alpha_1 m + \alpha_2 (m^+)^{-1} \det m
\edm
transforms in the same manner as $m$.
For a real, diagonal mass matrix, the transformation amounts to
\be \label{KM1}
m_u' = \alpha_1 m_u + \alpha_2 m_d m_s \hspace{3mm} (\mbox{cycl.}\; u
\rightarrow d \rightarrow s \rightarrow u)\fs\end{equation}
Symmetry does therefore not distinguish $m'$ from $m$. Since the effective
theory exclusively exploits the symmetry properties of QCD, the above
transformation of the quark mass matrix does not change the form of the
effective Lagrangian -- the transformation may be absorbed in a suitable
change of the effective coupling constants.\cite{Kaplan Manohar} This implies,
however, that the expressions obtained with this Lagrangian
for the masses of the pseudoscalars,
for the scattering amplitudes, as well as for the matrix elements of the
vector and
axial
currents are invariant under the operation
$m\!\rightarrow
\!m'$. Conversely, the experimental information on these observables
does
not allow us to distinguish $m'$ from $m$. 

Within the
approximations used, we may equally well write the mass ratio that 
characterizes the ellipse
in the form $Q^2\!=\!(m_s^2-\frac{1}{2}m_u^2-\frac{1}{2}m_d^2)/(m_d^2-m_u^2)$.
Up to terms of order $m^4$, which are beyond the accuracy of
our formulae,
the differences of the squares of the quark masses are invariant under
the transformation (\ref{KM1}). Hence $Q^2$ is invariant, i.e.\ the ellipse 
is mapped onto itself.
The position on the ellipse, however, does not remain invariant and can
therefore not be determined on the basis of chiral perturbation 
theory alone.

We are not dealing with a hidden symmetry of QCD here -- this
theory is not invariant under the change (\ref{KM1}) of the quark masses.
In particular, the matrix elements of the scalar and pseudoscalar operators
are modified. The
Ward identity for the axial current implies, for example, that
the vacuum--to--pion matrix element of the pseudoscalar density is given
by \be\langle 0|\bar{d}\,i\gamma_5 u|\pi^+\rangle=\sqrt{2}\,F_{\pi}
M_{\pi^+}^2/(m_u+m_d)\fs\end{equation}
The relation is exact, except for electroweak corrections. It involves the
physical quark masses and is not invariant under the above transformation.
Unfortunately, however, an experimental probe sensitive to the scalar or
pseudoscalar currents is not available --
the electromagnetic
and weak interactions happen to probe the low energy structure of the system
exclusively through
vector and axial currents.

\section{Estimates and bounds}\label{theory}

I now discuss the size of the corrections to the leading order formulae
(\ref{i3}) for the two quark mass ratios $m_u/m_d$ and
$m_s/m_d$. For the reasons just described, this discussion
necessarily involves a theoretical input of one sort or
another. To clearly identify the relevant ingredient, I explicitly formulate
it as hypothesis {\it A, B,} $\ldots$

\subsection*{Hypothesis A: Assume that the
corrections of order $m^2$ or higher are small and neglect
these. }

This is the attitude taken in early work on the
problem.\cite{Phys Rep} In the notation used above, the assumption
amounts to $\Delta_M\!\simeq\! 0$. In the plane spanned by
$m_u/m_d$ and $m_s/m_d$, this condition represents a
straight line, characterized by $m_s/\hat{m}\simeq
(2M_K^2-M_\pi^2)/M_\pi^2\simeq 26$.  The intersection with the ellipse then
fixes things. It is convenient to parametrize the
position on the ellipse by means of the ratio $R$ that measures the relative
size of isospin and SU(3) breaking,
\be R\equiv\frac{m_s-\hat{m}}{m_d-m_u}\fs\ee
With the value $Q\!=\!24.2$ (Dashen theorem), the intersection
occurs at the mass ratios given by Weinberg, which correspond to
$R\!\simeq\!43$. For the value of the semi--axis
which follows from
$\eta$ decay, $Q\!=\!22.7$, the intersection instead takes place at
$R\!\simeq\!39$.

The baryon octet offers a
good test: Applying the hypothesis to the chiral expansion of the
baryon masses, i.e. disregarding terms of order $m^2$,
we arrive at three independent estimates for $R$, viz.
$51\pm 10\;(N\!-\!P)$, $43\pm 4 \;(\Sigma^-\!-\!\Sigma^+)$ and
$42\pm 6\; (\Xi^-\!-\!\Xi^0)$.\footnote{Note that, in this case, the expansion
contains terms of order $m^{\!\frac{3}{2}}$, which do
play a significant role numerically. The error bars represent simple
rule--of--thumb estimates, indicated by the
noise visible in the calculation. For details see ref.~\cite{Phys Rep}.}
Within the errors, these results are consistent with the
values $R\simeq 43$ and $39$, obtained above from $K^0\!-\!K^+$ and
from $\eta\!\rightarrow\!\pi^+\pi^-\pi^0$, respectively. A recent
reanalysis of $\rho\!-\!\omega$ mixing~\cite{Urech1} leads to
$R\!=41\pm4$ and thus corroborates the picture further.

Another source of information concerning the ratio of isospin and SU(3)
breaking effects is the branching ratio
$\Gamma_{\psi'\rightarrow\psi \pi^0}/\Gamma_{\psi'\rightarrow\psi\eta}$.
The chiral expansion of the corresponding ratio of
transition amplitudes starts with: \cite{Ioffe Shifman}
\bdm \frac{\langle\psi\pi^0|\,\qbar m q|\psi'\rangle}
          {\langle\psi\eta|\,\qbar m q|\psi'\rangle}
=\frac{3\sqrt{3}}{4\,R}\{1+\Delta_{\psi'}+\ldots\}\fs\edm
Disregarding the correction $\Delta_{\psi'}$, which is
of order $m_s\!-\!\hat{m}$, the data
imply $R\!=\!31\pm4$, where the error bar corresponds to the experimental
accuracy
of the branching ratio. The value is significantly lower than those listed
above. The higher order corrections are discussed in
ref.~\cite{DW psi prim}, but the validity of the multipole
expansion used there is
questionable.\cite{Luty
Sundrum} The calculation is of interest, because it is independent
of other determinations, but at the present level of theoretical understanding,
it is subject to considerable uncertainties. Since the quark mass ratios
given in refs.~\cite{DHW mass ratios} rely on the value
of $R$ obtained in this way, they are subject to the same reservations.
Nevertheless, the information
extracted from $\psi'$ decays is useful, because it puts an upper limit on the
value of $R$. As an SU(3) breaking effect, the correction $\Delta_{\psi'}$ is
expected to be of order 25\%. The estimate
$|\Delta_{\psi'}|<0.4$
is on the conservative side. Expressed in terms of $R$, this implies $R<44$.

\subsection*{Hypothesis B:
Assume that the effective coupling constants are dominated
by the singularities which are closest to the origin.}

I have discussed this generalization of the
vector meson dominance hypothesis
in section \ref{pecc}. 
Since the coupling constant $L_5$ and the 
combination $L_5- 12 L_7 - 6
L_8$ are known experimentally (from $F_K/F_\pi$ and $3M_\eta^2+M_\pi^2-4M_K^2$,
respectively), we may express $\Delta_M$ in terms of known quantities,
except for a contribution from $L_7$.
Inserting the estimate (\ref{e16}) for this constant, we obtain a small, 
negative
number:~\cite{Leutwyler1990}
$\Delta_M\simeq -0.16$. Unfortunately, the result is
rather sensitive to the uncertainties of the saturation hypothesis,
because the contributions from $L_7$ and from the remainder are of opposite
sign and thus partly cancel. This is illustrated by the following
observation. The constant $L_7$ enters through its contribution to the mass
of the $\eta$. When replacing the term with the one from $\eta'$--exchange, the
corresponding energy denominator, $(M_{\eta'}^2-p^2)^{-1}$, should be 
evaluated at
$p^2\!=\!M_\eta^2$. In the
formula (\ref{e16}), the denominator is replaced by the
corresponding leading order term, $(M_{\eta'}^2)^{-1}$. The neglected
higher order effects are not exceedingly large, but they reduce the
numerical value of the prediction for $\Delta_M$ by a factor of 2.

The breaking of SU(3) induces $\eta\!-\!\eta'$ mixing.
When analyzing this effect
within chiral perturbation theory,\cite{GL SU(3)} we noticed that the
observed value of $M_\eta$ requires a mixing angle that is about twice as
large as the canonical value $|\mixingangle |\simeq 10^\circ$ accepted
at that time. The conclusion was confirmed experimentally soon
thereafter.\cite{Phenomenology}
We may now turn the argument around,\cite{Leutwyler1990} use the phenomenology 
of the mixing
angle to estimate the magnitude of $L_7$ and then determine the size of
$\Delta_M$. For a mixing angle in the range
$20^\circ<\mixingangle<25^\circ$, this leads
to $-0.06<\Delta_M<0.09$. In this calculation, the
energy denominator is evaluated at the proper momentum, but the uncertainties
arising from the cancellation of two contributions remain.

Quite irrespective of these uncertainties, the result
for $\Delta_M$ is a very small number: The hypothesis that the low energy
constant $L_7$ is dominated
by the singularity due to the $\eta'$ implies that the corrections to the
lowest order mass formula for $M_K^2/M_\pi^2$ are small. In view of the 
elliptic constraint, this amounts to the statement that
{\it A} follows from {\it B}.

\subsection*{Hypothesis C:
Assume that the large--$N_c$ expansion makes sense for
$N_c\!=\!3$.}

As noted already in ref.~\cite{Gerard}, the ambiguity discussed in section
\ref{KM} disappears in the large--$N_c$ limit, because the
Kaplan--Manohar transformation violates the Zweig rule. In this limit, the
structure of the effective theory is modified, because,
as briefly mentioned in section \ref{U(1)},
the U(1)--anomaly is then suppressed, so that the spectrum contains a ninth
Goldstone boson.\cite{Large Nc}
The implications for the effective Lagrangian are extensively discussed in
the literature and the leading terms in the
expansion in powers of $1/N_c$ are well--known.\cite{Leff U(3)}
More recently, the analysis was extended to first non--leading order,
accounting
for all terms which are suppressed either by one power of $1/N_c$ or by one
power of the quark mass matrix.\cite{bound}
This framework leads to a bound for $\Delta_M$, which
arises as
follows.

At leading order of the chiral expansion, the mass of
the $\eta$ is given by
the Gell--Mann--Okubo formula. At the next order of the expansion,
there are two categories of corrections:
(i) The first is of the same origin as the correction which occurs in the mass
formula (\ref{i5}) for the ratio $M_K^2/M_\pi^2$
and is also determined by $\Delta_M$. The expression for the
mass of the $\eta$, which follows from the Gell--Mann--Okubo formula,
$M_\eta^2\!=\!\frac{1}{3}(4M_K^2-M_\pi^2)$, is replaced
by \bdm m_1^2=\mbox{$\frac{1}{3}$}(4M_K^2-M_\pi^2)+\mbox{$\frac{4}{3}$}
(M_K^2-M_\pi^2)\,\Delta_M\fs\edm
(ii) In addition, there is mixing between the two states $\eta,\eta'$. The
levels repel in proportion to the square of the transition matrix
element $\sigma_1\!\propto\!\langle\eta'|\,\qbar m q |\eta\rangle$,
so that the mass formula for the $\eta$ takes the form
\be\label{mass formula}
M_\eta^2=m_1^2-\frac{\sigma_1^2}{M_{\eta'}^2-m_1^2}\fs\ee
This immediately implies the inequality $M_\eta^2\!<\!m_1^2$, i.e.
\bdm\Delta_M>-\frac{4M_K^2-3M_\eta^2-M_\pi^2}{4(M_K^2-M_\pi^2)}=-0.07\fs\edm

At leading order of the expansion, the transition matrix element $\sigma_1$ is
given by $\sigma_0=\frac{2}{3}\sqrt{2}\,(M_K^2-M_\pi^2)$. There are again two
corrections of first non--leading order:
$\sigma_1=\sigma_0\,(1+\Delta_M-\Delta_N)$. The first is an SU(3) breaking
effect of order $m_s-\hat{m}$, determined by $\Delta_M$, while $\Delta_N$
represents
a correction of order $1/N_c$ of unknown size --
the mass formula (\ref{mass formula}) merely fixes $\Delta_N$ as a function of
$\Delta_M$ or vice versa: As $\Delta_M$ grows,
$\Delta_N$
decreases.
A coherent picture, however, only results if both
$|\Delta_M|$ and $|\Delta_N|$ are small compared with unity.
If the above inequality were saturated,
$\sigma_1$ would have to vanish, i.e.
$1+\Delta_N-\Delta_M\!=\!0$. In other words, the corrections would have
to cancel the leading term. It is clear that,
in such a situation, the expansion is out of control. Accordingly,
$\Delta_M$ must be somewhat larger than $-0.07$. Even $\Delta_M\!=\!0$ calls
for large Zweig rule violations,
$\Delta_N\simeq\frac{1}{2}$.
The condition
\be\label{i8} \Delta_M\!>\!0\ee
thus represents a generous lower bound for
the region where a truncated $1/N_c$ expansion leads to meaningful results.
It states
that the current algebra formula, which relates the quark mass ratio
$m_s/\hat{m}$ to the meson mass ratio $M_K^2/M_\pi^2$, represents an upper
limit, $m_s/\hat{m}\!<\!2M_K^2/M_\pi^2-1\!=\!25.9$.

This shows that {\it A, B} and {\it C} are mutually consistent, provided
$\Delta_M$ is small and positive.
The bound (\ref{i8}) is shown in fig.\ 3: Mass ratios in the hatched region are
in conflict with the hypothesis that the first two terms of the $1/N_c$
expansion yield meaningful results for $N_c\!=\!3$. Since the
Weinberg ratios correspond to $\Delta_M\!=\!0$, they
are located at the boundary of this region. In view
of the elliptic constraint, the bound in particular implies
$m_u/m_d\,\raisebox{0.2em}{$>$}\hspace{-0.8em}
\raisebox{-0.3em}{$\sim$}\,\frac{1}{2}$.

\subsection*{Hypothesis D: Assume that $m_u$ vanishes.}

It is clear that this
assumption violates the large--$N_c$ bound just discussed. {\it D} is
also inconsistent with {\it A} and {\it B}. In fact, as pointed out in
refs.~\cite{Dallas Florida}, this hypothesis
leads to a very queer picture, for the following reason.

The lowest order mass
formulae (\ref{e3}) and (\ref{e4}) imply that the ratio $m_u/m_d$ determines
the $K^0/K^+$ mass difference, the scale being set by $M_\pi$:
\bdm
M^2_{K^0} - M^2_{K^+} = \frac{m_d - m_u}{m_u + m_d}\, M^2_\pi + \ldots
\edm
The formula holds up to corrections from higher order terms in the chiral
expansion and up to e.m.\ contributions. Setting $m_u\!=\!0$, the relation
predicts $M_{K^0}-M_{K^+}\simeq 16\,\mbox{MeV}$, four times larger than the
observed mass difference. The disaster can only be blamed on the higher
order terms, because the
electromagnetic self energies are much too small.
Under such circumstances,
it does not make sense to truncate the expansion at first non--leading order.
The conclusion to be drawn from the assumption $m_u=0$ is that
chiral perturbation theory is unable to account for the masses of the
Goldstone bosons. It is difficult to understand how a framework with a
basic flaw like this can be so successful.

The assumption $m_u\!=\!0$ also implies that
the
matrix elements of the scalar and pseudoscalar currents must exhibit very
strong SU(3) breaking effects.\cite{Dallas Florida} Consider
for instance  the pion and kaon matrix elements of the scalar
operators $\ubar u,\dbar
d, \sbar s$. In the limit $m_d=m_s$, the ratio
\bdm r =\frac{\langle\pi^+|\,\ubar u-\sbar s|\pi^+\rangle}
             {\langle K^+|\,\ubar u-\dbar d|K^+\rangle} \edm
is equal to 1. The SU(3) breaking effects are readily calculated by working
out the derivatives of $M_{\pi^+}^2,M_{K^+}^2$ with respect to $m_u,m_d,m_s$.
Neglecting the chiral logarithms which turn out to be small in this case, the
first order corrections may be expressed in terms of the masses,
\bdm r =
\left(\frac{m_s-m_u}{m_d-m_u}
\cdot\frac{M_{K^0}^2-M_{\pi^+}^2}{M_{K^0}^2-M_{K^+}^2}
\right)^{\!\raisebox{-0.2em}{{\small 2}}}
\left \{\rule{0em}{1.2em}1 + O(m^2)\right\}\fs\edm
The relation is of the same
character as the
one that leads to the elliptic constraint: The corrections
are of second order in the quark masses. For $m_u\!=\!0$, the elliptic
constraint reduces to $m_s/m_d\!=\!Q+\frac{1}{2}$, so that the relation
predicts \footnote{The precise value depends on the number used for the
electromagnetic
contribution to $M_{K^+}\!-\!M_{K^0}$.} $r\simeq
3$, while SU(3) implies $r\simeq 1$.
So, $m_u=0$ leads
to the prediction that the evaluation of the above matrix elements with sum
rule or lattice techniques will reveal extraordinarily strong flavour symmetry
breaking effects -- a bizarre picture. For me this is enough to
stop talking about $m_u\!=\!0$ here.

\section{Magnitude of $m_s$}
Finally, I briefly comment on the absolute magnitude of the quark masses.
The effective low energy theory does not allow us to determine these
phenomenologically, because the
low energy constant $B$ cannot be measured directly.
The best determinations of the magnitude of $m_s$ rely on
QCD sum rules.\cite{QCD SR1}
A detailed discussion of the method in application to the mass spectrum of
the quarks was given in 1982.\cite{Phys Rep}
The result for the $\overline{\mbox{MS}}$ running mass at scale
$\mu=1\,\mbox{GeV}$ quoted in that report is
$m_s=175\pm 55\,\mbox{MeV}$. The issue has been investigated in
considerable detail since then.\cite{QCD SR2} The value given in
the most recent paper,\cite{BPdR}\be m_s=175\pm 25 \,\mbox{MeV}\co\ee
summarizes the state of the art:
The central value is confirmed and the
error bar is reduced by about a factor of two.
The residual uncertainty
mainly reflects the systematic errors of the method, which
it is difficult to narrow down further.

There is considerable progress in the numerical simulation of QCD on a
lattice.\cite{Mackenzie} For gluodynamics and bound states of heavy
quarks, this approach already yields significant results. The values obtained
for $m_s$ are somewhat smaller than the one given above. The APE
collaboration,\cite{Martinelli} for instance, reports
$m_s=128\pm18\,\mbox{MeV}$ for
the $\overline{\mbox{MS}}$ running mass at $\mu\!=\!2\,\mbox{GeV}$. It is
difficult, however,
to properly account for the vacuum fluctuations generated by quarks with small
masses. Further progress with light
dynamical fermions is required before the numbers obtained for $m_u,m_d$ or
$m_s$ can be taken at face value. In the long run, however, this method
will allow an accurate determination of all of the quark masses.

\section{Conclusion}
\begin{figure}[t]
\centering
%\mbox{ \epsfbox{fitep3.eps} }
\mbox{\epsfysize=8cm \epsfbox{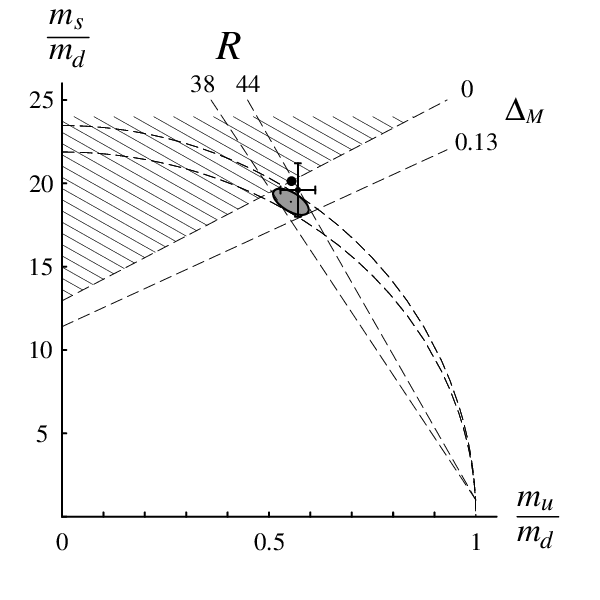} }
\parbox{11.9cm}{Figure 3: Quark mass ratios. The dot corresponds to
Weinberg's
values, while the cross represents the estimates given in ref.~\cite{Phys
Rep}. The hatched region is excluded by the bound $\Delta_M>0$. The
error ellipse shown is characterized by the constraints $Q=22.7\pm
0.8$, $\Delta_M>0$, $R<44$, which are indicated by dashed lines.} \end{figure}

Light quark effective theory represents a coherent theoretical framework
for the analysis of the low energy structure of QCD. The method has
been applied to quite a few matrix elements of physical interest and the
predictions so obtained have survived the experimental tests performed until
now. In these lectures, I focussed on the implications for the ratios of the
light quark masses, where the method leads to the following results:

1. The ratios $m_u/m_d$ and $m_s/m_d$ are constrained to an ellipse, whose
small semi--axis is equal to 1,
\bdm \left( \frac{m_u}{m_d}\right)^2+\frac{1}{Q^2}\,
     \left( \frac{m_s}{m_d}\right)^2=1\fs\edm
$\eta$ decay yields a remarkably
precise measurement of the large semi--axis,
\bdm Q=22.7\pm0.8\fs\edm
Unfortunately, however, the experimental situation concerning the lifetime of
the $\eta$ is not satisfactory -- the given
error bar relies on the averaging procedure used by the
Particle Data Group.\\
2. The position on the ellipse cannot accurately be determined from
phenomenology alone. The theoretical arguments given imply
that the corrections to
Weinberg's leading order mass formulae are small. In particular, there is a
new bound based on the $1/N_c$ expansion,
which requires
$m_u/m_d\,\raisebox{0.2em}{$>$}\hspace{-0.8em}
\raisebox{-0.3em}{$\sim$}\,\frac{1}{2}$ and thereby eliminates the
possibility that the $u$--quark is massless.\\
3. The final result for the
quark mass ratios is indicated by the shaded error ellipse in fig.\ 3, which is
defined by the following three constraints: (i) On the
upper and lower sides, the ellipse is bounded by the two dashed lines
that correspond
to $Q=22.7\pm0.8$. (ii) To the left, it touches the hatched region,
excluded by the large--$N_c$ bound. (iii) On the right, I use the upper limit
$R<44$, which follows from the observed value of the
branching ratio
$\Gamma_{\psi'\rightarrow \psi\pi^0}/
\Gamma_{\psi'\rightarrow \psi\eta}$.
The corresponding range of the various parameters of interest is~\cite{ratios}
\bea \frac{m_u}{m_d}=0.553\pm0.043\co\;\;\;
     \frac{m_s}{m_d}\al=\al18.9\pm0.8\co\;\;\;
     \frac{m_s}{m_u}=34.4\pm3.7\co\no
 \frac{m_s-\hat{m}}{m_d-m_u}= 40.8\pm 3.2\co\;\;\;
\frac{m_s}{\hat{m}}\al=\al 24.4\pm1.5\co\;\;\;
\Delta_M
= 0.065\pm0.065
\fs\nonumber\eea
While
the central value for $m_u/m_d$ happens to coincide with the leading
order formula, the one for $m_s/m_d$ turns out to be slightly smaller. The
difference, which amounts to
6\%, originates in the fact that the
available
data on the $\eta$ lifetime as well as the lattice result for
$(M_{K^+}\!-\!M_{K^0})_{e.m.}$ imply a somewhat smaller value of $Q$ than
what is predicted by the Dashen theorem.\\
4. The theoretical arguments discussed as hypotheses {\it A, B} and
{\it C} \,in section \ref{theory} are perfectly consistent with these numbers.
In particular, the early determinations of $R$,
based on the baryon mass splittings and on $\rho$--$\omega$ mixing,\cite{Phys
Rep} are confirmed. The rough estimate $m_s/\hat{m}\!=\!29\pm7
$, obtained by Bijnens, Prades and de Rafael from QCD sum
rules,\cite{BPdR} provides an independent check: The lower end of this
interval
corresponds to $\Delta_M< 0.17$. Fig.\ 3 shows that this constraint restricts
the allowed region to the right and is only slightly weaker than the condition
$R<44$ used above.\\
5. The mass of the strange quark is known quite accurately from QCD sum
rules:
\bdm m_s=175\pm25\,\mbox{MeV}\;\;\;(\overline{\mbox{MS}}\mbox{ scheme at}\;
\mu=1\,\mbox{GeV})\fs\edm
Using this value, the above ratios then determine the
size of $m_u$ and $m_d$:
\bdm m_u=5.1\pm 0.9\,\mbox{MeV}\co\;\;\;   m_d=9.3\pm\,1.4\,\mbox{MeV}\fs
\edm

\end{document}